\documentclass[11pt]{article}
\usepackage{moriond,epsfig}

\def\Journal#1#2#3#4{{#1} {\bf #2}, #3 (#4)}

\def\NIMA{{\em Nucl. Instrum. Methods} A}

\def\be{\begin{equation}}
\def\ee{\end{equation}}
\def\bea{\begin{eqnarray}}
\def\eea{\end{eqnarray}}

\begin{document}
\vspace*{4cm}
\title{START-UP OF THE NEMO 3 EXPERIMENT}

\author{ F. PIQUEMAL on behalf of the NEMO collaboration }

\address{Centre d'Etudes Nucl\'{e}aires de Bordeaux-Gradignan, \\ CNRS-IN2P3
 et Universit\'{e} de Bordeaux I, F-33175 Gradignan Cedex, FRANCE}

\maketitle\abstracts{The NEMO collaboration is looking to measure
 neutrinoless double beta decay down to a lower limit of 0.1 eV for the 
effective neutrino mass. The NEMO 3 detector is now operating in the 
Fr\'ejus Underground Laboratory. The expected performance and the data of the
first runs are presented here.}

\section{Introduction}

The recent discovery of neutrino oscillations is proof that the neutrino
is a massive particle. However, the oscillation experiments are only 
sensitive to the difference in the square of the masses of  two eigenstates 
of the neutrino.
One method for measuring the absolute mass scale of neutrinos is through
a careful investigation of the end point energy of single beta decay. Another method is  through neutrinoless double beta decay($\beta\beta(0\nu)$) which is the mission of the NEMO 3 detector. 
  This process has
 not yet been observed to date~\cite{ba}.
The $\beta\beta(0\nu)$ process is the decay of an (A,Z) nucleus to an
(A,Z+2) nucleus by simultaneous emission of two electrons without neutrino 
emission. The non-conservation of lepton number is a
 signature of physics beyond the Standard Model. Different ways are 
possible for this decay in the SU(2)$_L$ $\times$ U(1)$_Y$, SU(2)$_L$ 
$\times$ SU(2)$_R$ $\times$ U(1)$_Y$ or supersymmetric models \cite{Mo}. 
The simplest one is to have a light neutrino Majorana exchange in the
SU(2)$_L$ $\times$ U(1)$_Y$ model. In this case, the $\beta\beta(0\nu)$
  half-life depends on the effective neutrino mass.
In 1989, the NEMO (Neutrinoless Experiment with Molybdenum) 
 Collaboration\footnote{CEN-Bordeaux-Gradignan, France; CFR-Gif/Yvette, 
France; CRN-Strasbourg, France; Department of Physics-Jyvaskyla, Finland;
 FNSPE-Prague, Czesch. Republic; INEEL, Idaho Falls, USA; INR-Kiev, Ukraine; ITEP-Moscow, Russia; 
JINR-Dubna, Russia; LAL-Orsay, France; LPC-Caen, France; MHC-South Hadley, 
USA; Saga University, Japan.} started a R\&D program to build a detector which would be able to study 
 the effective neutrino mass down to about 0.1 eV by looking for the
$\beta \beta$0$\nu$ decay process. The NEMO 3 detector  is now operating
 in the Fr\'ejus Underground Laboratory (4800 m.w.e.). A brief description and
 the expected performance of the detector are presented in
 this article as well as the first events.

\section{The NEMO 3 detector}
\subsection{Description}

\begin{figure}
\begin{center}\psfig{figure=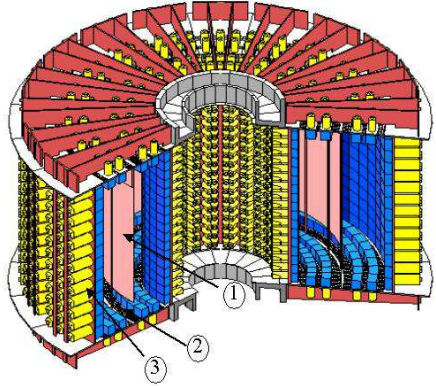,height=2.5in}\end{center}
\caption{Drawing of the NEMO 3 detector: 1. Source foil (up to 10 kg), 2.
 tracking volume (6180  Geiger cells) 3.  Calorimeter (1940 plastic 
scintillators coupled to low activity photomultipliers).
\label{fig:nemo3}}
\end{figure}

The detector is cylindrical in design and divided into
20 equal sectors (Fig.~\ref{fig:nemo3}). A thin (40-60 mg/cm$^2$) cylindrical 
source foil of $\beta\beta$ emitters has been constructed from 
either a metal film or powder bound by an organic glue to mylar strips.
The detector houses up to 10 kg of these isotopes.

The source hangs between
two concentric cylindrical tracking volumes consisting of open octagonal
drift cells operating in Geiger mode. These cells
run vertically and are staged in a 4, 2, and 3 row pattern
to optimize track reconstruction. The design of the  drift cells calls for  
50 $\mu$m  anode and cathode wires to prevent rapid aging. The tracking volume 
is filled with a mixture of 96\% helium + 4\% ethanol. The radial and
 longitudinal resolution are 0.5 mm (1 $\sigma$) and 1 cm (1 $\sigma$), respectively. 
Consequently, the precision of the position of the emission vertex  of the two electrons is
0.6 and 1.8 cm in the transverse and longitudinal directions, respectively.

The external walls of these
tracking volumes are covered by calorimeters made of large blocks of plastic
 scintillator coupled to very low radioactivity 
3" and 5" Hammamatsu PMTs. The energy resolution depends on the scintillator
 shape and the associated PMT. It  ranges from 11\% to
14.5\% (FWHM) for 1 MeV electrons. The time resolution is 250 ps (1 $\sigma$)
at 1 MeV. A laser calibration system permits daily checks on the stability of the
energy and time calibration parameters.
The detector contains 6180 Geiger cells and 1940 scintillators.

Additionally, a solenoid surrounds the detector and produces  a field of 30 Gauss to reject pair production events. Finally, external shielding first in the
form of 20 cm of low activity iron reduces the gamma ray flux and 
then 30 cm of water suppresses the flux of  neutrons.
All the materials used in the detector were selected for their high radiopurity by $\gamma$-ray spectroscopy via Germanium detectors. 

The detector is able to identify electrons, positrons, $\gamma$-rays and 
delayed-$\alpha$ particles. This allows the detector to measure the internal contamination of the source by electron-$\gamma$, electron-$\gamma$ 
$\gamma$ or electron-$\gamma$-$\alpha$ channels and to reject the
external background.

\subsection{The sources}

Several sources were placed in the detector which not only study $\beta \beta($0$\nu$) 
decay but also measure other process such as the $\beta \beta($2$\nu$) decay, the
Majoron decay mode and the external backgrounds.
Table~\ref{tab:iso}
 summarizes the list of the isotopes with their total mass and decay modes of interest. Note that the $^{100}$Mo, $^{82}$Se and $^{116}$Cd
isotopes  search for $\beta \beta($0$\nu$) decay, 
$\beta \beta($2$\nu$) decay to the ground  and excited states, and Majoron emission
 decay $\beta \beta($0$\chi$). 

The other enriched isotopes ($^{130}$Te, $^{150}$Nd,
$^{96}$Zr and  $^{48}$Ca) were installed to measure the 
$\beta \beta($2$\nu$) half-life for comparison with the predictions of
different nuclear matrix element calculations. Another interest in these sources is to 
measure the contamination
after enrichement, chemical purification and source foil  production for
 future improvements in NEMO 3. For $^{100}$Mo, 10$^5$  
$\beta \beta($2$\nu$) events per year will be recorded giving high statistics
 for the angular distribution between the two emitted electrons and the single 
electron energy spectrum.  

The natural tellurium and copper are very pure, so the events on these sources
in the 3 MeV region are only induced by the external $\gamma$-rays flux.

\begin{table}[t]
\caption{List of enriched isotopes placed in the NEMO 3 detector 
.\label{tab:iso}}
\vspace{0.4cm}
\begin{center}
\begin{tabular}{|c|c|c|}
\hline
Isotope & Mass (g) & Intended Studies \\
\hline
 $^{100}$Mo & 7200 &  $\beta \beta($0$\nu$), $\beta \beta($2$\nu$),  
$\beta \beta($0$\chi$) \\
 $^{82}$Se & 1000 &  $\beta \beta($0$\nu$), $\beta \beta($2$\nu$),  
$\beta \beta($0$\chi$) \\
 $^{116}$Cd & 600 &  $\beta \beta($0$\nu$), $\beta \beta($2$\nu$),  
$\beta \beta($0$\chi$) \\
 $^{130}$Te & 1300 &   $\beta \beta($2$\nu$) \\
 $^{150}$Nd & 48 &   $\beta \beta($2$\nu$) \\
 $^{96}$Zr & 20 &   $\beta \beta($2$\nu$) \\
 $^{48}$Ca & 10 &   $\beta \beta($2$\nu$) \\
 $^{nat}$Te & 800 &  External background \\
 Cu & 600 &  External background \\ 
\hline
\end{tabular}
\end{center}
\end{table}

For the $^{100}$Mo and $^{82}$Se isotopes, the  $\beta \beta($0$\nu$) 
signal is expected to be centered around 3 MeV ( 3.034 and 2.993 MeV, respectively). In this 
energy region the two electron background can be divided in two components:

- the internal background is produced by the decay of  
$^{208}$Tl (Q$_{\beta}$=4.99 MeV) and $^{214}$Bi (Q$_{\beta}$=3.27 MeV) and by the high energy tail of the 
 $\beta \beta($2$\nu$) spectrum given the energy resolution of NEMO 3.

- the external background is produced by interactions of high energy $\gamma$-rays 
created by radiative capture of neutrons inside the detector or by muons 
 via bremstrahlung.

After chemical or physical purification, the contamination levels of  
$^{208}$Tl and $^{214}$Bi were measured by $\gamma$-ray
spectroscopy. The contamination levels for the $^{100}$Mo and $^{82}$Se sources are
given in Table~\ref{tab:pur}. Only limits have been obtained for $^{100}$Mo.
In the case of $^{82}$Se, this source was measured in the NEMO 2 
detector and  the contamination was found to correspond to ``hot spots''
 which can be removed by cuts in the data related to the position of an event's vertex.

\begin{table}[t]
\caption{Measured values of the contamination from $^{208}$Tl and $^{214}$Bi in
the $^{100}$Mo and $^{82}$Se sources in mBq/kg.\label{tab:pur}}
\vspace{0.4cm}
\begin{center}
\begin{tabular}{|c|c|c|}
\hline
Isotope & $^{208}$Tl & $^{214}$Bi \\
\hline
 $^{100}$Mo & $<$ 0.02 & $<$ 0.3\\
 $^{82}$Se & 0.4 $\pm$ 0.1 & 1.2 $\pm$ 0.5\\
\hline
\end{tabular}
\end{center}
\end{table}

\subsection{First data with 20 sectors}

\begin{figure}
\begin{center}\psfig{figure=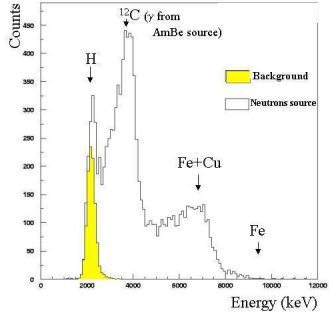,height=2.5in}\end{center}
\caption{Spectrum of one crossing  electron events with an Am-Be neutron source.
\label{fig:neut}}
\end{figure}

In March of 2002 the data acquisition started without shielding to measure 
external backgounds and to obtain data with a neutron source. The most sensitive channel~\cite{ma} to see the effect of neutrons inside the detector is the one-crossing-electron channel corresponding to Compton electrons created in a 
 scintillator and then crossing the detector. This kind of event
is distinguished from the two electron events emitted from the source by 
time-of-flight measurements.  Fig.~\ref{fig:neut} shows the recorded spectrum of 
crossing electrons with a neutron source and no shielding.
The main contribution of $\gamma$-rays comes from radiative captures of 
 neutrons in the copper and iron and  the 4.43 MeV $\gamma$-rays
 emitted by the Am-Be neutron source. The Compton peak at 1.8 MeV corresponds to the
  2.2 MeV $\gamma$-rays from neutron capture on hydrogen. The grey region, represents
 the normalized background for crossing  electron events without the neutron source,
 it will be suppressed  by a factor of 100  by a 20 cm iron shield. 

\section{Prospect and conclusion}

The number of background events has been calculated by a Monte-Carlo simulation
using the GEANT 3.21 CERN program. The contribution of the internal background is
less than 1 event per year for the $^{100}$Mo and the $^{82}$Se sources in the
 [2.8-3.2] MeV region. In the case of $^{82}$Se, if the ``hot spots'' are 
rejected, the internal background is effectively zero events because the
 $\beta \beta($2$\nu$) half-life is 10 times longer than that for $^{100}$Mo.
 In five years, no events are expected from the external background~\cite{ma}.
Thus, in five years the sensitivity in terms of limits at the 90 \% confidence level for
the effective neutrino mass is $\langle m_\nu \rangle <$ 0.2 - 0.7 eV and 
 $\langle m_\nu \rangle <$ 0.6 - 1.2 eV for 7.2 kg of  $^{100}$Mo and 1 kg of
 $^{82}$Se, respectively. The range in the neutrino mass is due to the 
uncertainties in the nuclear matrix elements.

Data collection with a full shield (iron + neutron shielding) will start this
summer.

\end{document}